\documentclass[
	a4paper, 
	10pt, 
	unnumberedsections, 
	twoside, 
]{LTJournalArticle}
\usepackage[inkscapelatex=false]{svg}
\usepackage[pdf]{graphviz}
\usepackage{xpatch}
\makeatletter
\newcommand*{\addFileDependency}[1]{
  \typeout{(#1)}
  \@addtofilelist{#1}
  \IfFileExists{#1}{}{\typeout{No file #1.}}
}
\makeatother
\xpretocmd{\digraph}{\addFileDependency{#2.dot}}{}{}
\usepackage{subcaption}
\usepackage{fancyvrb,fancybox}

\newenvironment{FramedVerb}
  {\VerbatimEnvironment
   \begin{Sbox}
   \begin{minipage}{\linewidth}
   \begin{Verbatim}}
  {\end{Verbatim}
   \end{minipage}
   \end{Sbox}
   \fbox{\TheSbox}}

\runninghead{} 

\footertext{\textit{RISC-V Summit Europe, Bologna, 8-12 June 2026}} 

\title{Reproducibility in open-source RISC-V HW flows} 

\author{%
 Peter Kourzanov\textsuperscript{1}\thanks{Corresponding author: \href{mailto:peter.kourzanov@imec.be}{\tt peter.kourzanov@imec.be}}
 \and
 Anmol Anmol\textsuperscript{1}
}

\date{
\vspace{-.5em}
\footnotesize\textsuperscript{\textbf{1}}imec DSRD CSA, {\em Kapeldreef 75, 3001 Leuven, Belgium}
\vspace{-1.5em}
}

\renewcommand{\maketitlehookd}{%
\begin{abstract}
\vspace{-.5em}
\noindent Open-source hardware is booming. To prevent fragmentation, encourage collaboration and reuse we propose the RISC-V to join forces with Reproducible Builds communities and concentrate innovation potential where its needed most: creation of new micro-architectures, IPs and their integration into new SoCs and applications. To facilitate this goal, we chose Guix, a rigorous solution for reproducible software artefacts. We apply it to dependency management \& reproducibility problems in open-source hardware and show the validity of the approach, taking CVA6 as a running example. The end result - a fully reproducible collection of packaged tests, emulation, simulation and cycle-accurate models - shows a promising workflow that could (in future) scale to support larger RISC-V community with reusable software \& hardware components for next-generation platforms.
\end{abstract}
\vspace{-1.5em}
}

\begin{document}

\maketitle 

\section{Introduction and related work}
\vspace{-.5em}

Open-source hardware is booming. Repositories of openly available IPs such as OpenCores, PULP, as well as frameworks similar to Chipyard, ESP, OpenPiton, BlackParrot, OpenTitan, X-heep, next to eco-systems like SpinalHDL and Litex as well as organizations such as lowRISC, ChipsAlliance and OpenHW Group see growing adoption and contributions.

While this is a positive trend, diversification is not always an improvement taken for granted. Let's consider the (open-source) \emph{software} eco-systems, which have experienced strong growth from the early days of UNIX, POSIX and the C/C++ programming languages, inspiring free implementations such as BSD, GCC and the Linux kernel and eventually enabling the rise of Linux open-source stack (i.e., LAMP) to the widespread commercial adoption. There, standardizing common design philosophies, OS A\{P,B\}I and PL sides of this success story have shown how effective can community be in improving the ensuing \emph{software} eco-system, and of course the RISC-V community's aim is to repeat this success story for the emerging \emph{hardware} eco-systems. There's one catch, though.

The ease with which software components used to be developed and modified comes with relaxed requirements on reproducibility. The development cycles are short and are forgiving to trial-and-error methodology. For hardware development, nothing like that is true:
during design, implementation and fabrication stages, every successive step is ever more resource intensive.

\vspace{-1.5em}
\section{Methodology}
\vspace{-.5em}

We identify several bottlenecks that need to be addressed in order for RISC-V eco-system to thrive especially in competitive advanced technology nodes:

\vspace{-.5em}
\begin{description}
\item[architectural] issues arising when innovating micro-architecture for new SoCs. {\em Gem5 design-space exploration full-system simulations take too long.}
\item[implementation] difficulties of obtaining realistic circuits for open-source micro-architectures, primarily due to the lack of SoTA EDA tooling. But also: \underline{\em verification \& fine-tuning synthesis takes time.}
\item[back-end] challenges of mapping circuits to competitive PDKs and interconnect structures. {\em Back-end design takes a long time to optimize, debug and sign-off. Reuse and knowledge sharing is essential}
\end{description}

\vspace{-1em}
\noindent To ease verification and stimulate collaboration we propose 3 essential techniques borrowed from \cite{BuildSystemsAlaCarte}: \emph{Dependency management}, \emph{Minimalism}, \emph{Early cutoff \& Self tracking}. By adopting these ideas for open-source hardware flows the RISC-V community can accelerate innovation and throughput, ensuring success of R\&D.

\vspace{-.5em}
\begin{figure}[h]
\includesvg[width=.5\textwidth]{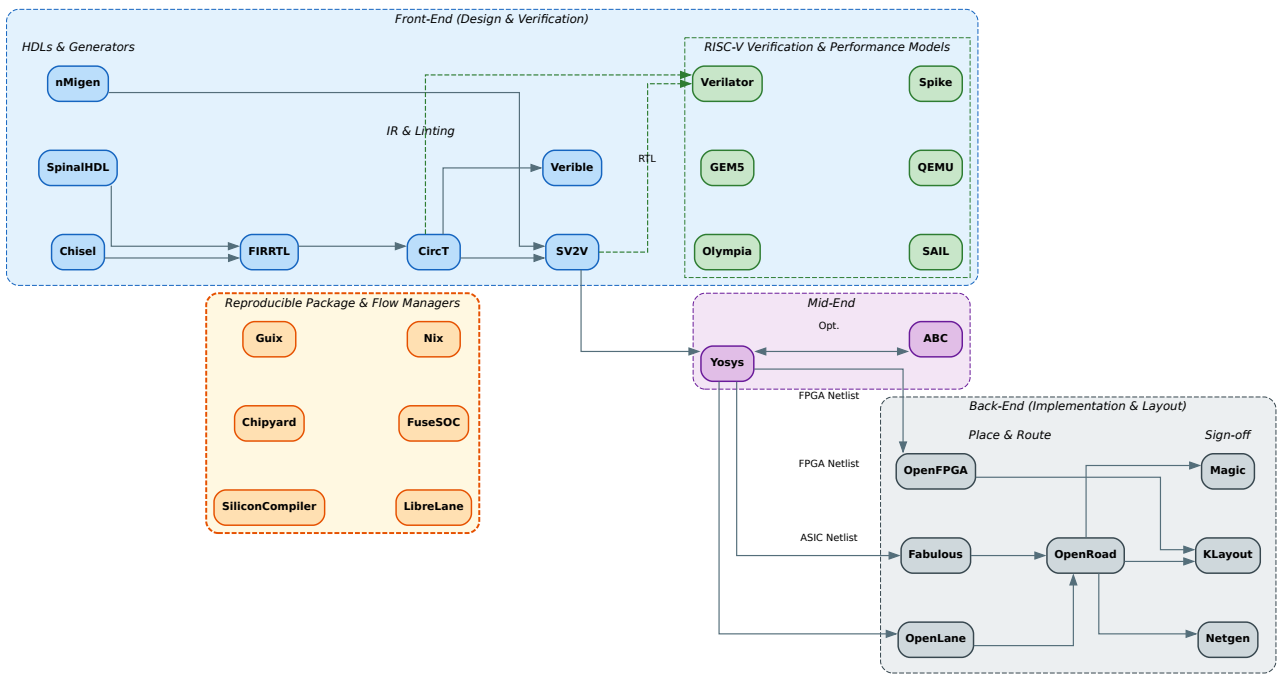}
\caption{Plethora of open-source EDA tools \& flows}
\label{fig:tools}
\end{figure}

\vspace{-3em}
\subsection{Dependency management}
\vspace{-.5em}

Just like software build systems, hardware flows require consistent maintenance of dependencies: every step of the flow needs to use a specific version of the tool (i.e., one from Fig.\ref{fig:tools}) for reproducibility. Dependencies need to be complete and parsimonious with no noise.

\vspace{-.5em}
\subsubsection{Determinism}

is essential for reproducibility, hardware flow products need determinism to scale. Engineers from across the world must be able to replicate bit-exact artefact outputs (just like reproducible SW).

\vspace{-.5em}
\subsubsection{Static \& Dynamic}

(i.e., computed) dependencies must be both supported by reproducible hardware flows. Although some existing frameworks such as Chipyard already use a software build tool (GNU make) to manage the hardware simulation, synthesis and mapping flows, the lack of support for dynamic dependencies in GNU make is an impediment.

\subsection{Early cutoff \& Self-tracking}
\vspace{-.5em}

During development many aspects of hardware flows need to undergo modification, including HDL source code, documentation collaterals, etc. Sometimes the build recipes themselves are modified, i.e., the hardware flow should support \emph{self-tracking}. Of course, when one changes comments in the source file, the artefact does not necessarily need to be fully rebuilt, so the hardware flow should also support \emph{early cutoff}. 

\vspace{-1.5em}
\subsection{Minimalism}
\vspace{-.5em}

The paper above defines this for build tools \emph{``executing tasks at most once per build
and only if they transitively depend on inputs changed since the previous build.''} \cite{BuildSystemsAlaCarte}

\vspace{-.5em}
\subsubsection{Precise hashing}

To accomplish caching, support determinism, self-tracking and early cutoff a build system may opt for a thorough application of \emph{precise hashing} whereby dependencies, environment, build recipes and artefact outputs are all hashed. The hashes of products of the hardware flow serve as the evidence of build recipe's success in mapping inputs to outputs.

\vspace{-.5em}
\subsubsection{Cloud-readiness}

Last but not least, the open-source hardware flow must be readily to be deployable to the cloud, enabling multiple groups of innovators to collaborate and share products in a reproducible way.

\vspace{-1em}
\begin{figure}[h]
\includesvg[width=.5\textwidth]{riscv-test-flows.svg}
\caption{Guix dependency graph for \texttt{riscv-test-flows}}
\label{fig:packages}
\end{figure}

\vspace{-2.5em}
\section{Experiment with OpenHW CVA6}
\vspace{-.5em}

To demonstrate the proposed flow - following \cite{Batten22guix} - we created a set of new Guix packages encapsulating 5 sub-flows, CVA6 and \verb|riscv-tests|, reused existing Guix packages for relevant tools: Spike \cite{Spike}, Qemu \cite{Qemu}
, Verilator
, with the related cross-compilation toolchains and other supporting Guix packages as dependencies.

\vspace{-.5em}
\begin{figure}[h]
{\scriptsize
\begin{FramedVerb}
$guix time-machine -C channels.scm -- build -L . riscv-test-flows
/gnu/store/wkxya9aqd3r2...1v46831ccn26r2-riscv-test-flows-1.0.0
\end{FramedVerb}
}
% ...
\caption{Standardized build \& test process}
\label{fig:build}
\end{figure}

\vspace{-1em}
These packages are depicted in Fig.\ref{fig:packages} as \emph{rectangles} for the standard and new customized tools for the RISC-V architecture that we're using (\verb|rv64imafdc|) and as \emph{ovals} for new flow-related packages. The entry-point is the \verb|riscv-test-flows| which validates \verb|riscv-tests| on the following sub-flows: \verb|qemu-flow|, \verb|spike-flow|, \verb|vp-flow|, \verb|gem5-flow| and \verb|cva6-flow|. 
Each sub-flow depends on the relevant \{emu,simu\}lator tool (\verb|qemu|, \verb|spike|, \verb|vp| \cite{vp-plus-plus}, \verb|gem5| \cite{gem5} and \verb|verilated-cva6|, respectively) and is responsible for running all \verb|riscv-test|s and for storing the complete report in the artefact output. In our approach the entry-point package obtains access to these reports from the sub-flow dependencies and fails the build if unexpected errors are detected.

\vspace{-1.75em}
\section{Conclusions}
\vspace{-1em}

Once Guix knows where to find the packages (in Fig.\ref{fig:build} above, after a clone is checked-out from our github repo \cite{guix-cva6}), standard Guix build commands initiate and manage all the aspects mentioned earlier: deterministic dependency management, early cutoff, self-tracking, minimalism and publishing/pulling available artefacts to/from other machines. The final result is an empty artefact in \verb|/gnu/store| - with a precise hash - which witnesses fully reproducible verification.

\vspace{-.5em}
\subsubsection{``Why not Nix''} is a fair question, as both Guix and Nix share same heritage. While many open hardware developers are now considering Nix as a "better" package manager for hardware, we do see advantages that Guix offers: a powerful and standard recipe implementation language (Scheme), full bootstrap (improving security, certifiability) and having a great community.

\vspace{-.5em}
\subsubsection{Future work}
Linking this work to more \{emu,simu\}lators and RTL frameworks as well as extending the approach to cover back-end flows next to verification is an obvious next step. One nice future exploration would be to link this work to \cite{Ravanan} which could further streamline proliferation of reproducible hardware flows. Finally, deployment of this approach across sites and organizations using Guix substitute servers would provide a great boost to the speed of RISC-V adoption across both academic \& industrial R\&D institutions in Europe and elsewhere.

\vspace{-1.5em}
{\footnotesize
\printbibliography 
}

\end{document}